\documentclass[aps,prl,twocolumn,superscriptaddress,showpacs,nofootinbib,preprintnumbers,10pt,floatfix]{revtex4-2}

\usepackage{graphicx}
\usepackage{hyperref}
\usepackage[dvipsnames]{xcolor}
\usepackage{amsmath,amsthm,amssymb}
\usepackage{comment}

\renewcommand{\Im}{\mathrm{Im}\,}

\begin{document}
\preprint{LA-UR-23-29892}
\preprint{INT-PUB-23-037}

\title{Mitigating a discrete sign problem with extreme learning machines}

\author{Scott Lawrence}
\email{srlawrence@lanl.gov}
\affiliation{Department of Physics, University of Colorado, Boulder, CO 80309, USA}
\affiliation{Los Alamos National Laboratory Theoretical Division T-2, Los Alamos, NM 87545, USA}
\author{Yukari Yamauchi}
\email{yyama122@uw.edu}
\affiliation{Institute for Nuclear Theory, University of Washington, Seattle, WA 98195, USA}

\date{\today}

\begin{abstract}
An extreme learning machine is a neural network in which only the weights in the last layer are changed during training; for such networks training can be performed efficiently and deterministically. We use an extreme learning machine to construct a control variate that tames the sign problem in the classical Ising model at imaginary external magnetic field. Using this control variate, we directly compute the partition function at imaginary magnetic field in two and three dimensions, yielding information on the positions of Lee-Yang zeros.
\end{abstract}

\maketitle

In seminal papers by Lee and Yang~\cite{PhysRev.87.404,PhysRev.87.410}, phase transitions in many-body systems are investigated by studying the location of zeros of the partition function in the complex plane of a control parameter, such as the temperature or an external field. For example, the temperature at which a phase transition occurs can be extracted by examining the location of the zeros as the thermodynamic limit is taken. Given such a connection between the locations of zeros and properties of phase transitions, the location of the Lee-Yang zeros has been studied in a variety of systems in both theory~\cite{PhysRevLett.84.4794,PhysRevE.65.036110,PhysRevLett.110.248101, PhysRevE.97.012115, PhysRevB.92.125132, PhysRevE.96.032116, Krasnytska_2016, deger2020lee,Connelly:2020gwa,Rennecke:2022ohx,Johnson:2022cqv, Clarke:2023noy} and experiment~\cite{PhysRevLett.81.5644,PhysRevLett.109.185701,PhysRevLett.114.010601,PhysRevLett.118.180601}. Of particular interest to this Letter, the location of the Lee-Yang zeros of the classical Ising model has been studied by the high cumulants of thermodynamic observables~\cite{PhysRevE.97.012115, deger2020lee}, tensor networks~\cite{PhysRevB.92.125132}, and on a complete and random graph~\cite{Krasnytska_2016}. 

Direct computation of the Ising partition function at imaginary external field is obstructed by the presence of a sign problem, broadly similar to those that prevent the simulation of real-time quantum dynamics via the path integral and the computation of the nuclear equation of state at finite fermion density. A great collection of methods has been developed over the past few decades to treat sign problems that occur in lattice simulations; among them complex Langevin~\cite{Aarts:2008rr}, the density of states method~\cite{Langfeld:2016mct}, canonical methods~\cite{Alexandru:2005ix,deForcrand:2006ec}, reweighting methods~\cite{Fodor:2001au}, series expansions in the chemical potential~\cite{Allton:2002zi}, fermion bags~\cite{Chandrasekharan:2013rpa}, and analytic continuation from imaginary chemical potentials~\cite{deForcrand:2006pv}, and contour deformation methods~\cite{Alexandru:2020wrj}. Contour deformations methods in particular have recently been combined with machine learning approaches~\cite{Alexandru:2017czx,Ohnishi:2019ljc,Kashiwa:2019lkv,Alexandru:2018fqp,Alexandru:2018ddf,Lawrence:2021izu} including for theories with complex couplings~\cite{Lawrence:2022afv}; unfortunately, these methods are unable to treat systems with discrete degrees of freedom.

In this Letter we introduce a machine learning approach for treating a lattice sign problem, inspired by previous work on control variates~\cite{Lawrence:2020kyw,Lawrence:2022dba,Bhattacharya:2023pxx,Bedaque:2023ovz}, which does not depend on the ability to analytically continue the action and observables. (See~\cite{Kashiwa:2023dfx} for another approach to generalizing contour deformation methods to spin systems.) We demonstrate the method in studying the Lee-Yang zeros of the classical Ising model on a lattice $\Lambda$, whose partition function is given by
\begin{equation}
    Z(J;h) = \sum_{s \in \{0,1\}^{|\Lambda|}} \exp\Bigg\{J \sum_{\langle x,y\rangle} s_x s_y + h \sum_x s_x\Bigg\}
    \text,
\end{equation}
where the first sum in the exponential is taken over all pairs of neighboring lattice sites. We will use both a two-dimensional square lattice and a three-dimensional cubic lattice, each with periodic boundary conditions.

The partition function is polynomial in fugacity $e^{h}$, and at fixed $J$ is therefore determined up to a constant factor by the location of its zeros. The Lee-Yang theorem states that all zeros lie on the imaginary $h$ axis, on which the partition function is purely real. At larger volumes the zeros become more dense as a function of $-i h$, and when $J$ corresponds to the second-order phase transition this line of zeros reaches the real axis in the infinite-volume limit.

To find the zeros, we will operate at fixed $J$ and compute the ratio
\begin{equation}\label{eq:ratio}
    \frac{Z(h;J)}{Z_Q(J)} = \bigg\langle \exp\Big\{ h \sum_x s_x\Big\}\bigg\rangle_Q \equiv \langle \mathcal O \rangle_Q\text,
\end{equation}
where the ``quenched'' expectation value of $\mathcal O$ is computed relative to the Boltzmann distribution at $h=0$. This expectation value has a signal-to-noise problem related to the configuration-dependent phase of $\mathcal O \equiv e^{h \sum s}$: the variance of the phase is always of order unity, while the signal, as a ratio of partition functions, falls exponentially with the volume.

We increase the signal-to-noise ratio by constructing an appropriate control variate, as was done in~\cite{fernandez2009mean,weigel2010error,Bhattacharya:2023pxx} for other signal-to-noise problems, and in~\cite{Lawrence:2020kyw,Lawrence:2022dba} for various sign problems. We will find a function $f(s)$---termed the \emph{control variate}---with vanishing expectation value, so that $\langle \mathcal O \rangle = \langle \mathcal O - f \rangle$. The variance of the new observable is
\begin{equation}
    \mathrm{Var}(\mathcal O - f) = \langle \mathcal O^2\rangle - 2\langle \mathcal O f\rangle + \langle f^2\rangle + \langle \mathcal O \rangle^2
    \text.
\end{equation}
Thus when $f$ is strongly correlated with $\mathcal O$, the new function $\tilde {\mathcal O} \equiv \mathcal O - f$ has a smaller variance than the original observable $\mathcal O$. We will refer to $\tilde{\mathcal O}$ as the variance-reduced observable.

To avoid introducing any difficult-to-control bias in the Monte Carlo calculation, we will construct $f(s)$ in such a way as to guarantee that its expectation value vanishes exactly. Defining a discrete differentiation operator by
\begin{equation}
    \nabla_x g(\vec s) \equiv g(s) - g(s)|_{s_x\rightarrow -s_x}
    \text,
\end{equation}
we begin with a function $g(s)$ and construct the control variate $f(s)$ according to
\begin{equation}\label{eq:transdiff}
    f(s) e^{-S} \equiv \sum_x \nabla_{x=x_0} g(T_x s)
    \text.
\end{equation}
Here the sum is taken over all lattice sites $x$ and $T_x$ is the translation operator. The most general translationally invariant control variate $f(s)$ can be expressed in this way---translational invariance is desirable in this case as the observable of interest, $\mathcal O = e^{-i h \sum s}$, is also translationally invariant. The choice of differentiation site $x_0$ is irrelevant due to the sum over translations.

Any choice of $g(s)$ will yield a valid control variate, but not all will improve the signal-to-noise ratio. As in~\cite{Bhattacharya:2023pxx}, we begin the optimization process by noting that given a basis of candidate control variates $F_i$, the optimal linear combination may be determined by measuring the correlations $M_{ij} = \langle F_i F_j\rangle$ and $v_i = \langle \mathcal O F_i\rangle$, and computing
\begin{equation}
    c = M^{-1} v\text.
\end{equation}
The optimal control variate (within the chosen basis) is now given by $f = \sum_i c_i F_i$.

To improve on the performance of a directly constructed basis of control variates, we take inspiration from \emph{extreme learning machines}~\cite{huang2006extreme}. An extreme learning machine (henceforth ELM) is a neural network in which only the final layer is trained; all other weights are left equal to their pseudorandomly initialized values. The learning process is now linear regression, which is both efficient and deterministic. The loss in expressivity from the fact that most weights are fixed, is at least partially compensated by the ability to have a far larger network for the same computational cost.

\begin{figure}
    \centering\includegraphics[width=0.95\linewidth]{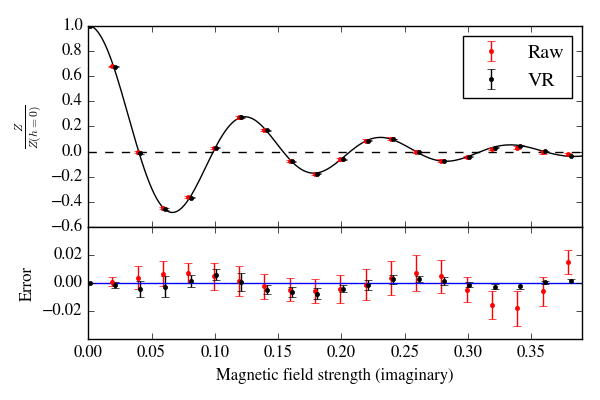}
    \caption{Ratio of partition functions on an $8\times 8$ lattice at $J=0.4$.  The bottom panel shows the deviation from the exact result, given by the transfer matrix method.
    \label{fig:verification}}
\end{figure}

In this Letter we define $g(s)$ via an ELM with a single hidden layer. The inputs are $N$ functions $h_j(s)$, detailed below. The hidden layer is of width $W L^d$, where $L^d$ is the spacetime volume of the lattice and $W$ is a tuneable width scaling factor. We use the CELU function~\cite{barron2017continuously} for the activation layer. Thus the ELM can be written in the form
\begin{equation}
    g(x) = c_i \sigma_{\mathrm{CELU}}(A_{ij} x_j)
    \text.
\end{equation}
Only the parameters in the vector $c$ are optimized. The $N \times W L^d$ matrix $A$ is left in its randomly initialized state for the entire procedure: each component of $A$ is drawn independently from the uniform distribution on the interval $[-L^{-d},L^{-d}]$. As an implementation detail, the differentiation and averaging defined by Eq.~(\ref{eq:transdiff}) are performed before multiplication by $c$. This allows the optimization of the coefficients $c$ to be performed by directly solving a linear system, just as in~\cite{Bhattacharya:2023pxx}.

In principle the spin configuration might be directly used as an input to the ELM. In practice, as is often the case in machine learning tasks, a large boost in performance is seen when the inputs are augmented with hand-crafted functions of the spin configuration. Here, we select inputs to the ELM by trial and error.

First, from the spin configuration $s$ we construct a `scaled' version
\begin{equation}
    \tilde s_x = e^{-D(x)/D_0} s_x
    \text,
\end{equation}
where the distance function $D(x)$ is a measure of the distance from $x$ to the origin:
\begin{equation}
    D(x) = \left[\sum_{k=1}^d 2 - 2 \cos\left(\frac{2\pi x_k}{L}\right)\right]^{1/2}\text.
\end{equation}
This construction has the effect of encouraging the ELM to focus on short-distance physics. We take $D_0 = 0.5$ throughout.

The input to the ELM consists of a total of $2 + 3 L^d$ elements. The scaled spin configuration $\tilde s$ accounts for $L^d$ of those. We also include the real and imaginary parts of the phase of the Boltzmann factor (that is, $\cos \Im S(s)$ and $\sin \Im S(s)$). Finally, we include two scaled copies of $\tilde s$: $\tilde s \cos \Im S(s)$ and $\tilde s \sin \Im S(s)$.

The detailed training procedure is as follows. The weights of the ELM are independently drawn from the uniform distribution specified above. We collect $K$ samples from the Ising model Boltzmann factor at some $J$ (but $h=0$), and these samples are split into two sets. The first set, of size $K_{\mathrm{learn}}$, is used only in fitting the optimal weights of the ELM, while the second (of size $K_{\mathrm{est}} = K - K_{\mathrm{learn}}$) is used for evaluating expectation values. This separation is necessary to avoid bias in the final computed expectation values. Throughout this letter the two sets will be chosen to be of equal size.

\begin{figure}
    \centering\includegraphics[width=0.95\linewidth]{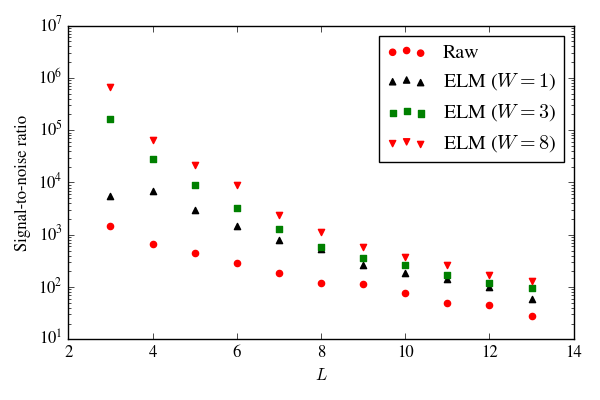}
    \caption{The performance of control variates constructed from an ELM across lattice sizes and ELM widths, as measured by the ratio of the magnitude of the expectation value of $\mathcal O$ to the variance of the estimator. All calculations are performed at $J=0.2$ and an external field of $h=0.1i$, with $5 \times 10^4$ samples used to train each ELM and $5 \times 10^4$ samples per data point.\label{fig:scaling}}
\end{figure}

On each sample the ELM gives $W L^d$ outputs, which we will name $g_i(s)$. Each output is differentiated with respect to the origin according to the finite differencing operator defined above and summed over possible translations, defining a basis $f_i(s)$ of possible control variates. The correlations $M_{ij}$ and $c_i$ are measured on the $K_{\mathrm{learn}}$ samples, and from those measured values the optimal coefficients $c_i$ estimated. This defines the control variate to be used, and the improved observable
\begin{equation}
    \tilde{\mathcal O} \equiv \mathcal O - \sum_i c_i f_i
\end{equation}
is measured on the remaining samples.

One additional technical detail must be treated: the correlation matrix $M$ is typically ill-conditioned, with a condition number that rises rapidly with the width parameter $W$. We regularize $M$ by adding a small multiple of the identity matrix ($10^{-10}$ in this Letter).

To verify the correctness of this method, we first work with the model in two dimensions. At modest values of $L$, the partition function may be computed exactly by means of the transfer matrix. Three calculations of the partition function on an $8\times8$ lattice are shown in Fig.~\ref{fig:verification}. The data points are the Monte Carlo estimate of the ratio in Eq.~(\ref{eq:ratio}) with or without the variance reduction method applied. Each calculation is done with $K_{\mathrm{learn}} = 5 \times 10^3 = K_{\mathrm{est}}$ samples, and a width scaling of $W = 3$. Both data agree with the exact result, while the errors from the calculation with the variance reduction are seen to be substantially smaller than those without the reduction. The coupling of $J=0.4$ is chosen to be slightly hotter than the critical coupling in two dimensions of $J_c \approx 0.441$~\cite{PhysRev.65.117}.

Fig~\ref{fig:scaling} shows the performance of the ELM, measured by the variance of the estimator, as a function of the size of the lattice. We select a high temperature and weak magnetic field, of $J=0.2$ and $h=0.1$, to avoid zeros of the partition function and make this ratio meaningful. In additional to the unimproved estimator, two computations are shown, corresponding to ELM widths of $L^2, 3 L^2,$ and $8 L^2$. We see that for any fixed size of ELM, there is no exponential improvement in the variance, only a factor which is fixed or decaying as $L$ increases. The typical improvement seen, a factor of $\sim 10$, corresponds to an advantage of $10^2$ in computational time when high precision is desired.

\begin{figure}
    \centering\includegraphics[width=0.95\linewidth]{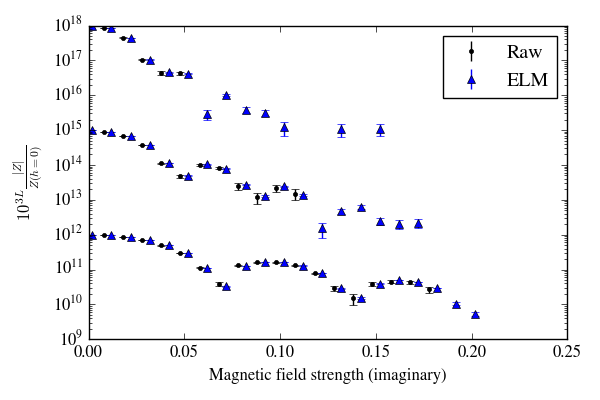}
    \caption{The partition function of the Ising model, as a function of complex magnetic field strength, for different volumes in three spacetime dimensions ($4^3$, $5^3$, and $6^3$). The increasing density of the zeros at larger volumes is clearly visible. At larger volumes and magnetic field strengths, the raw data is insufficient to distinguish the partition function from $0$, making it impossible to localize zeros further from the real axis; the use of an ELM improves the situation somewhat, enabling the location of zeros to be further constrained. Data points within two standard deviations of $0$ are hidden for readability.\label{fig:three}}
\end{figure}

Finally, in Fig.~\ref{fig:three} we show the partition function at imaginary magnetic field on a three-dimensional lattice, at lattice sizes of $L=4,5,6$. The largest lattice size, $6^3$, is beyond what can be computed via the transfer matrix with reasonable computational resources.  We take $J=0.2$ to again be slightly hotter than the phase transition (which sits at $J_c \approx 0.22$~\cite{talapov1996magnetization}). Each calculation is done with $K_{\mathrm{learn}} = 2 \times 10^4 = K_{\mathrm{est}}$ samples, and a width scaling of $W = 5$. 

We have detailed a practical algorithm for mitigating volume-scaling sign problems in lattice field theory. In the context in which we have tested the method---the Ising model at imaginary external magnetic field---it consistently yields a speedup of two orders of magnitude over the naive approach. This approach is directly applicable to spin systems and other models in which the degrees of freedom in the path integral are discrete, a marked advantage over previous machine learning approaches that make use of contour deformations; however, at this stage we have not attained an exponential improvement in the average phase.  This deficiency may be expected to be a fruitful direction for future work.

All results in this Letter make use of the JAX library Equinox~\cite{kidger2021equinox} for the implementation of the ELM. S.L.~is grateful to Frederic Koehler for originally suggesting extreme learning as an technique of interest.

S.L.~was supported at the beginning of this work by the U.S.~Department of Energy under Contract No.~DE-SC0017905, and subsequently by a Richard P.~Feynman fellowship from the LANL LDRD program. LANL is operated by Triad National Security, LLC, for the National Nuclear Security Administration of U.S. Department of Energy (Contract No.~89233218CNA000001). Y.Y.~is supported by the INT's U.S. Department of Energy grant No.~DE-FG02-00ER41132.

\bibliographystyle{apsrev4-2}
\bibliography{refs}

\end{document}